\documentclass[11pt]{article}

\usepackage{jcapmod}
\usepackage{booktabs}
\usepackage[english]{babel}
\usepackage{amsmath,amssymb,amsbsy,amstext, amsthm}
\usepackage{bm}
\usepackage{graphicx}
\usepackage{amsfonts}
\usepackage{amssymb}
 \usepackage{exscale,relsize}
 \usepackage[makeroom]{cancel}
\usepackage{soul}
\usepackage{bm}
\RequirePackage{color}

\usepackage{colortbl}
\definecolor{rp}{cmyk}{0.2, 1, 0.6, 0}
\definecolor{green2}{cmyk}{0, 1, 0.5, 0}
\definecolor{lightgreen}{cmyk}{0.2, 0, 0.2, 0.2}
\definecolor{lightgray}{cmyk}{0.1,0.2,0,0.1}
\definecolor{lightgray2}{cmyk}{0.4,0.4,0,0.8}
\definecolor{black}{cmyk}{1.0,1.0,1.0,1.0}

\allowdisplaybreaks[1]

\usepackage{colortbl}
\definecolor{lightgreen}{cmyk}{0.2, 0, 0.2, 0.2}
\definecolor{lightgray}{cmyk}{0.1,0.2,0,0.1}
\definecolor{lightgray2}{cmyk}{0.1,0.1,0,0.1}

\makeatletter
\newlength{\apb@width}
\newcommand{\autoparbox}[2][c]{\settowidth{\apb@width}{#2}\parbox[#1]{\apb@width}{#2}}

\makeatother

\numberwithin{equation}{section}

\def\be{\begin{equation}}
\def\ee{\end{equation}}

\def\bea{\begin{eqnarray}}
\def\eea{\end{eqnarray}}

\def\beq{\begin{equation}}
\def\eeq{\end{equation}}
\def\bea{\begin{eqnarray}}
\def\eea{\end{eqnarray}}

\def\0{{\boldsymbol 0}}

\DeclareRobustCommand{\SkipTocEntry}[4]{}

\begin{document}

\begin{titlepage}

\setcounter{page}{1} \baselineskip=15.5pt \thispagestyle{empty}
 \begin{flushright}\end{flushright}
\bigskip\

\vspace{1cm}
\begin{center}
    
{\fontsize{19}{28}\selectfont \sffamily\bfseries
Anisotropic tensor power spectrum at interferometer scales
induced by tensor squeezed non-Gaussianity }
\end{center}

\vspace{0.2cm}

\begin{center}
{\fontsize{14}{30}\selectfont   Angelo Ricciardone$^1$, 
Gianmassimo Tasinato$^2$}
\end{center}

\begin{center}
\textsl{$^1$ Faculty of Science and Technology, University of Stavanger, 4036, Stavanger, Norway}
\end{center}
\begin{center}
\textsl{$^2$ Department of Physics, Swansea University,
Swansea, SA2 8PP, U.K.}
\end{center}

\vspace{1.2cm}
\hrule \vspace{0.3cm}
\noindent 
{\sffamily \bfseries 
Abstract
} \\[0.1cm]
We develop a scenario of inflation with spontaneously broken time and space diffeomorphisms, with distinctive features for the
 primordial tensor modes.  Inflationary tensor fluctuations are non adiabatic, and can acquire a mass  during the inflationary epoch. 
 They can evade the Higuchi bound around  de Sitter space, thanks to interactions with the fields driving expansion. Correspondingly, the primordial stochastic gravitational wave background (SGWB) is characterised by a tuneable scale dependence, and can be detectable at interferometer scales. In this set-up, tensor non-Gaussianity can be parametrically enhanced in the squeezed limit. This induces a coupling between  long and  short tensor modes,  leading to a specific quadrupolar anisotropy in the primordial SGWB spectrum, which can be used to build estimators for tensor non-Gaussianity. We analyse  how our inflationary system can be tested with interferometers, also discussing  how an  interferometer  can be sensitive to a primordial anisotropic SGWB.   
\vskip 10pt
\hrule

\vspace{0.6cm}
 \end{titlepage}

 \tableofcontents


\vspace{1cm}

\section{Introduction}
\label{sec:intro}

 Cosmological inflation predicts the existence of a stochastic background 
 of gravitational waves (SGWB), whose power spectrum amplitude depends on the value of the Hubble parameter
 during inflation, as well as on the details of the inflationary process: see~\cite{Maggiore:1999vm,Guzzetti:2016mkm} for  reviews. In most inflationary models,
 the tensor power spectrum is almost scale invariant, with a slightly red tensor tilt. 
  In this case, the most
 promising way to detect the primordial SGWB is through Cosmic Microwave Background (CMB) B-mode polarisation (see e.g.~\cite{Kamionkowski:2015yta}).  There are however
 examples of inflationary scenarios where the amplitude of tensor modes can be amplified at scales much smaller
 than CMB's (see e.g.~\cite{Cook:2011hg,Cook:2013xea,Barnaby:2011qe,Biagetti:2014asa,Domcke:2016bkh}). These frameworks 
 are observationally interesting,  since they open the possibility for 
  gravitational wave interferometers to probe the primordial  SGWB, and to constrain
   inflationary models; see e.g.~\cite{Bartolo:2016ami} for a  study on the capabilities of the future Laser Interferometer Space Antenna (LISA)
 in this respect, containing the relevant references to the original literature. 
 
 Among the different scenarios which show interesting features for gravitational waves (GW) at small scales, here 
 we focus on a generalisation of solid inflation~\cite{Endlich:2012pz,Gruzinov:2004ty} dubbed supersolid inflation~\cite{Nicolis:2013lma,Bartolo:2015qvr}.
 This scenario explores the possibility of breaking both time and space diffeomorphisms during the inflationary
  epoch,  by means of space-dependent {\it vacuum expectation values} ({\it vevs}) of the  set of scalar fields that 
  drive  inflation. 
   Such a symmetry breaking pattern makes the
    primordial
   tensor modes non adiabatic: this implies that  tensor fluctuations acquire a  mass during inflation, and this affects the tensor power spectrum~\cite{Cannone:2014uqa}. 
  In the original  set-up~\cite{Endlich:2012pz}, a blue tensor spectrum was also found, with positive tensor spectral tilt proportional to the inflationary
  slow-roll parameter. 
 
  In this work we show that, within the context of supersolid inflation, it is possible to build concrete models with  parametrically larger positive values for the tensor spectral
 tilt. We can obtain this feature  by including appropriate non-minimal couplings between gravity and the
scalar fields driving inflation. One of the first results is that we evade the Higuchi bound~\cite{Higuchi:1986py,Fasiello:2013woa} in a de Sitter space, since the time and space-dependent $vevs$ of the scalar fields break 
 the de Sitter isometries that are needed for the applicability of Higuchi theorem.   A blue spectrum for tensor modes leads to a tensor power spectrum which increases at small scales, and the amplitude of SGWB can be
  sufficiently large to be  detectable at scales probed by interferometer. We discuss how LISA can probe a region of 
  parameter space  of  supersolid inflation models, and test their predictions for the amplitude of the  inflationary tensor spectrum.
  
  In~\cite{Ricciardone:2016lym} we shown that supersolid inflation can lead to  parametrically large tensor non-Gaussianity (nG) enhanced
  in the squeezed limit. The symmetry breaking pattern we analyse make the tensors non-adiabatic, and Maldacena's
  consistency relation on the squeezed limit of the tensor three-point (3-pt) function~\cite{Maldacena:2011nz} can be violated.  In this work, we study  how 
a large squeezed tensor nG leads to a quadrupolar anisotropy in the tensor power spectrum, induced
 by the coupling between long and short (wavelength) tensor modes. We discuss how 
the amplitude of the anisotropic contribution to the tensor power spectrum  can be used to build estimators for tensor non-Gaussianity. 
  We also start to analyse  the consequences of an anisotropic primordial SGWB for interferometers, and in particular we analyse  how the response function 
  of the instrument behaves in presence of such an anisotropic signal.\\
  
 The structure of the paper is the following: Section~\ref{sec:setup} discusses the supersolid set-up under consideration and the symmetries that dictate such a system. In Section~\ref{sec:backsol} we study the background evolution in two steps. We first analyze the dynamics of gravity non-minimally coupled to a single time-dependent scalar field; then we consider both time and space-dependent background fields and we  study   the features 
  of corresponding primordial cosmological fluctuations. In Section \ref{sec:pheno} we  analyse phenomenological implications of our findings and we compute the response of an interferometer to a primordial anisotropic SGWB. 
We conclude in Section~\ref{sec:disc} where we summarize our results and discuss possible future developments.  In Appendix~\ref{appendix}
we present  the equations for the Arnowitt-Deser-Misner (ADM) constraints~\cite{Arnowitt:1962hi}.

\section{Set-up and symmetries}
\label{sec:setup}

 \subsubsection*{The action}

We analyse a system whose field content and interactions   
 lead to a 
 space-time
symmetry breaking pattern 
with  novel features for  the inflationary tensor spectrum.

 We  are interested in a cosmological framework where both time
and space reparameterization symmetries are spontaneously broken by vacuum expectation values
of scalar fields~\cite{Nicolis:2013lma,Bartolo:2015qvr}.  Such system is a generalization of solid inflation~\cite{Endlich:2012pz}, where three derivatively coupled scalar fields have space-dependent
background values.
 For definiteness, we take as cosmological background de Sitter space-time; although 
it does not represent a  realistic inflationary set-up (inflation should end at some time, while de Sitter inflationary
expansion does not stop),
 nevertheless it is convenient for our discussion, which focuses 
 specifically on the dynamics of primordial tensor modes. 

 Our field content includes gravity and a quartet of scalar fields, $\phi$ and $\sigma^I$ ($I\,=\,1,2,3$), responsible for 
 breaking  space-time diffeomorphisms.  As we shall  discuss, our action is  motivated by symmetry considerations, and it reads as
 \be
 \label{eq:supaction}
S\,=\,\int\,d^4 x\,\sqrt{-g}\,\left[ {\cal L}_R-{\cal L}_{\phi}-M_{Pl}^4\,{\cal L}_{\sigma}\right]\,,
\ee
with
\bea
{\cal L}_R&=& \frac{M_{Pl}^2}{2}\,R\,, \label{eq:mainlag1}
\\
{\cal L}_{\phi}&=&\frac12 \,\partial_\mu \phi \partial^\mu \phi
+V_0+\frac{q_{\phi}}{M_{Pl}^2}\,G^{\mu\nu}\,\partial_\mu \phi \partial_\nu \phi \,, \label{eq:mainlag2}
\\
{\cal L}_{\sigma}&=&{q_A}\,F(\phi)\,{\text{Tr}}
\left(B\right)
+
{q_B}\,F^2(\phi)\,{\text{Tr}}
\left(B^2\right)\
+\lambda_0^2\,
{q_C}\,\frac{{\text{tr}}
\left(B^2\right)}{\left(\text{Tr} B\right)^2}
+
\frac{q_{\sigma}}{M_{Pl}^2}\,F(\phi)\,\delta_{IJ}\,\partial_\mu \sigma^I\,
\partial_\nu \sigma^J\,G^{\mu\nu}\label{eq:mainlag3}\,,
\eea

\noindent
where  $G^{\mu\nu}$ is the Einstein tensor, the Greek indices $\mu,\nu=\,0,\,1,\,2,\,3$
 denote spacetime coordinates,  $V_0$ is a constant potential, $q_i$ and $\lambda_0$ are dimensionless parameters and $M_{Pl}$ is the Planck mass. The
 scalar field $\phi$ has dimension of a mass, while the fields $\sigma^I$ have dimension of inverse of a mass.   
With {\rm ``Tr''} we indicate the trace of the field matrix $B_{IJ}$ which is defined as 
\be
B_{IJ}\,\equiv\,\partial_\mu \sigma_I \partial^\mu \sigma_J\,,\quad\quad\quad I,J=1,2,3\,.
\ee
We choose the function $F(\phi)$ as
 \bea
 \label{eq:effephi}
  F(\phi)&=&\exp{\left[  
2 H_0 \,\phi/\kappa_0^2
  \right]}\,,
  \eea
 where the two positive parameters $H_0$ and $\kappa_0$ have dimension of a mass. These two parameters
 play a special role in characterising the background configurations.  The scalar field
 $\phi$  acquires a time-dependent background profile $\phi(t)$ which
 spontaneously break time reparameterization \cite{Cheung:2007st}; the  triplet of scalars $\sigma^I$ have
 space dependent $vevs$ breaking space reparameterization. Interactions of $\phi$ and $\sigma^I$ 
 are described by the Lagrangians ${\cal L}_\phi$ and ${\cal L}_\sigma$  which include non-minimal couplings
 with gravity, described by operators proportional to $G_{\mu\nu} \,\partial^{\mu} \phi\,\partial^{\nu} \phi$
 and  $G_{\mu\nu} \,\partial^{\mu} \sigma^I\,\partial^{\nu} \sigma^J\,\delta_{IJ}$. Such non-linear derivative interactions belong
 to the family of Horndeski  scalar-tensor theories~\cite{Horndeski:1974wa}, and have been  first applied to inflation in~\cite{Kobayashi:2010cm,Germani:2010gm}. They are ghost
 free and do not lead to Ostrogradsky instabilities.
 We should  {\it not}  think to  them  as small corrections to the leading two derivative operators controlled by the kinetic terms. Instead we exploit their
 structure and focus on  branches of cosmological background configurations where their effects are particularly   relevant.  
 
 Action \eqref{eq:supaction} enriches the system examined in \cite{Ricciardone:2016lym} by including the operators $G^{\mu\nu} \partial_\mu \phi \,\partial_\nu \phi$, as well as 
  operators  quadratic in $B_{IJ}$. As we shall see, these contributions are important for characterizing the evolution of tensor modes during 
 inflation, in particular to avoid the Higuchi bound, to get a parametrically large positive tensor tilt, and enhanced tensor non-Gaussianity in the squeezed limit, which induces a quadrupolar anisotropy in the power 
 spectrum.

 \subsubsection*{Symmetries}

The structure of action \eqref{eq:supaction} is dictated by symmetry considerations, which we discuss here.
  We only consider covariant operators with up to four derivatives. In section
 \ref{sec:backsol}, we analyse background solutions of field
 equations which spontaneously break space-time isometries through $vevs$ for scalar fields. On the other hand, we recover
 these space-time  symmetries at the background level  by imposing internal shift and rotational symmetries for the fields. We impose
a shift  internal symmetry for each of the four scalars involved:
 \bea
 \phi&\to&\phi+c^0\,,
 \\
 \sigma^I&\to&\sigma^I+c^I\,,
 \eea
 where $c^0$,  $c^I$ are constants. Moreover, the three scalars $\sigma^I$ satisfy an internal rotational symmetry
 \be
 \label{eq:rotsymm}
 \sigma^I\to\Lambda^I_{\,\,\,J}\,\sigma^J\,,
 \ee
 where the constant matrix $\Lambda^I_{\,\,\,J} \in$ $SO(3)$, similar to the case of  solid inflation~\cite{Endlich:2012pz}.  Such internal symmetries allow one to recover the space-time symmetries spontaneuously 
 broken by  scalar $vevs$, and ensure isotropy and homogeneity for the geometrical background in accordance with observational CMB evidences~\cite{Ade:2015hxq}. 
 
 Additionally, we also impose an internal  scaling symmetry, distinctive of a scenario of supersolid inflation \cite{Domenech:2017kno,Ricciardone:2016lym}
  \bea
  \label{eq:scalsymm}
 \sigma^I&\to& \ell\,\sigma^I\,,
 \\
\phi &\to& \phi -\frac{\kappa_0^2}{H_0} \ln {(\ell)}
\,,
 \eea
for some constant  scaling parameter $\ell$. This scaling symmetry constrains the possible four-derivative  
 operators involving the scalars   $\sigma^I$ to the ones appearing in action \eqref{eq:supaction}, and motivates
the structure of the exponential coupling functions between $\phi$ and the scalar triplet $\sigma^I$. 
 The scaling symmetry \eqref{eq:scalsymm}, together with rotational symmetry \eqref{eq:rotsymm}, are  relevant for characterizing 
 properties  of the fluctuations \cite{Domenech:2017kno}, as we will discuss in the next Section.

\section{Background solutions and  cosmological fluctuations}
\label{sec:backsol}

We consider background configurations where the scalar fields spontaneously break time and space reparameterizations due to non-vanishing {\it vevs}.
We  focus on a de Sitter geometry for   the background, a space-time of sufficient generality  to analyse features of  primordial tensor modes: 
\be
\label{eq:desitter}
d s^2\,=\,-\,d t^2+e^{2 H_0 t}\,d \vec{ x\,}^2
\,,
\ee 
where $H_{0}$ is the Hubble parameter during the de Sitter phase. It is instructive to break space-time isometries in steps, so to appreciate the consequences of 
each class of operators contained in  ${\cal L}_\phi$ and ${\cal L}_\sigma$. 

\subsection{First step: spontaneous breaking of time reparameterization}
\label{subsec:timebreak}

We start considering a ``reduced'' action with respect to \eqref{eq:supaction}, only describing the dynamics of gravity non-minimally coupled with a 
single scalar field $\phi$
\be
S\,=\,\int d^4x\,\sqrt{-g}\,\left[ {\cal L}_R-{\cal L}_\phi\right]\,,
\ee
with ${\cal L}_R$, ${\cal L}_\phi$ given in \eqref{eq:mainlag1} and \eqref{eq:mainlag2} respectively.  Such action is an example of quartic Horndenski Lagrangian. 
 Applications to cosmological inflation of Horndenski-type couplings, as the ones above, have been much studied in the literature, starting with \cite{Kobayashi:2010cm,Germani:2010gm,Burrage:2010cu,Copeland:2012qf}. 

The background geometry is de Sitter space, \eqref{eq:desitter}, while we take the following homogeneous time-dependent ansatz for the scalar $\phi$
\be
\phi\,=\,\kappa_0^2\,t\,,
\ee
with $\kappa_0$ the same parameter appearing in the the function $F(\phi)$ in eq \eqref{eq:effephi}. Notice that if $\kappa_0$ is non-vanishing, then the scalar $vev$ spontaneously breaks 
time reparameterization: $\phi(t+\Delta t)\neq \phi(t)$. 

 Due to non-minimal coupling with gravity and the non-linearity of the equations, the system admits  two distinct branches of de Sitter solutions, given by
\begin{itemize}
\item {\bf Branch 1}
\bea
H_0^2&=& \frac{V_0}{3 M_{Pl}^2}\,,
\\
\kappa_0&=&0\,.
\eea
\item  {\bf Branch 2}
\bea
H_0^2&=& \frac{M_{Pl}^2}{6 q_{\phi}}\,, \label{conH0q2}
\\
\kappa_0^4&=&\frac{2 q_{\phi} V_0-M_{Pl}^4}{2 q_{\phi}}\,. \label{resk0a}
\eea
\end{itemize}
We are interested on the second branch, where the rate of expansion is inversely proportional to
 the parameter $q_{\phi}$  controlling  the non-minimal coupling between  the scalar field and gravity.
Parameterising the constant potential $V_0$ in terms of a quantity $v_1$ as
\be
V_0\,=\,3 H_0^2 M_{Pl}^2+2\,M_{Pl}^2\,v_1^2\,,
\ee
and using \eqref{conH0q2}, then eq \eqref{resk0a} becomes simpler
\be
\kappa_0^4\,=\,2 M_{Pl}^2\,v_1^2\,.
\ee
Hence the parameter $v_1$ controls the breaking of time-reparameterisation in this system. If $v_1=0$, then $\kappa_0=0$, and the two branches of solutions
coincide.   

\bigskip

The dynamics of cosmological fluctuations around the background solution  depends on the symmetry breaking parameter
 $v_1$. We analyse the quadratic actions for 
scalar and tensor fluctuations following closely the   methods we  implemented
in~\cite{Ricciardone:2016lym}, to which we refer the reader for more details.  We choose
an unitary gauge for scalar fluctuations, where the scalar $\phi$ is unperturbed,  while the metric fluctuations read
 (see also \cite{Mukhanov:1990me})
  \bea
  \label{eq:metric}
d s^2\,=\,
-\left(1+2 N\right)\,
 d t^2+2\,e^{2\, H_0\,t}\,\partial_i B\,d x^i dt+e^{2\, H_0\,t}\,\left(1+2 A \right)\,\delta_{ij}\,d x^i d x^j\,.
 \eea
The equations of motion for the scalar quantities $N$ and $B$, which are not dynamical, lead to the lapse and shift constraints~\cite{Maldacena:2002vr}. After imposing these constraint conditions, we find
  the following quadratic action for the scalar fluctuation $A$, the only scalar mode that  propagates:
 \bea
S_{A}^{(2)}\,=\,\left[
\frac{4\,v_1^4\,\left( v_1^2+3 H_0^2\right)}{9 H_0^2 \left( v_1^2+ H_0^2\right)^2}\right]
\,\times \int d t\,d^3 x\, a^{3}\left( 
\dot{A}^2-\frac{c_A^2}{a^2}
\left(\nabla A\right)^2
\right)\,,
\eea
where $c_A$ is the sound speed of the scalar fluctuation $A$
\be
c_A^2\,=\,\frac{v_1^2+H_0^2}{v_1^2+3 H_0^2}\,.
\ee
The overall coefficient in $S_A$ is proportional to a power of $v_1$,  controlling the time dependent $vev$
 of the scalar field profile.
  The scalar fluctuation $A$ can be thought as the Goldstone boson associated with the spontaneous breaking of time reparameterization,
and acquires non-trivial dynamics when $v_1\neq 0$, and hence the  scalar profile $\phi(t)$ is turned on. 
 
 \smallskip
 When we pass to consider the tensor sector it is convenient to parameterize it as \cite{Maldacena:2002vr}
 \be
 d s^2\,=\,-d t^2+a^2(t) \left(\delta_{ij}+\hat \gamma_{ij}\right)\,d x^i d x^j\,,
 \ee
 where $a(t)$ is the scale factor during the de Sitter phase, and $\hat{\gamma}$ can be expanded as
\be \label{exp-gam}
\hat{\gamma}_{ij}\,=\,\gamma_{ij}+\frac12 \gamma_{i k}\,\gamma_{k j}+\frac16 \gamma_{i k} \gamma_{k l} \gamma_{lj}\,,
\ee
where $\gamma_{ij}$ represents a first order, transverse ($\partial_{i} \gamma^{i}_{\,j}=0$) and traceless ($\gamma^{i}_{\,i}=0$) tensor fluctuation.
 The quadratic action for tensor modes around the de Sitter background configuration results
\be
S_{\rm \gamma}^{(2)}\,=\,\frac{M_{Pl}^2 \left(v_1^2+3 H_0^2 \right)}{12 H_0^2}\,\int d t\, d^3 x\,a^3\,\left( \dot{\gamma}_{ij}-\frac{c_T^2}{a^2}\left( \nabla \gamma_{ij} \right)^2  \right)\,,
\label{tens-act1}
\ee
where $c_T$ represents the tensor sound speed
\be \label{tenss1}
c_T^2\,=\,\frac{3 H_0^2-v_1^2}{3 H_0^2+v_1^2}\,.
\ee
Action \eqref{tens-act1} depends on the symmetry breaking parameter $v_1$, 
 both for what respect the effective Planck mass, which gets ``renormalised'' (see the coefficient in front
of eq \eqref{tens-act1}), and the tensor sound speed, see eq \eqref{tenss1}. When $v_1 \,=\,0$, one recovers the standard General Relativity (GR) result
for tensor fluctuations around de Sitter space~\cite{Riotto:2002yw}.   

The breaking of time reparameterization --   and 
 the direct coupling between  tensor fluctuations and  the fields driving inflation  --  lead to a distinct behaviour for
  tensor fluctuations in this  set-up, even around a pure de Sitter geometry \footnote{
   Notice that the tensor sound speed might be set to one through
a disformal transformation~\cite{Creminelli:2014wna,Baumann:2015xxa,Fumagalli:2016afy}, but this does not modify any physical consequences of the system since physical
 quantities do not depend on the frame~\cite{Baumann:2015xxa}. We will  not need to  consider such disformal transformation in the present  context.}.
 More drastic 
 consequences occur to  tensor perturbations   when breaking also space isometries, as we are going to discuss in the next subsection.
 
\subsection{Second step: spontaneous breaking of time and space reparameterizations}
\label{subsec:timespacebreak}

We proceed by considering a more general set-up, spontaneously breaking all space-time isometries
 by the $vevs$ of scalar fields, including also space reparameterizations.
 The total action that we now analyse  
 includes the fields $\sigma^I$, and reads as in eq \eqref{eq:supaction}.
The background profiles for $\phi$ and $\sigma^I$ which solve the background equations of motion are
\bea
\phi&=& \kappa_0^2\,t\,,
\\
\sigma^I&=& \lambda_0\,x^I\,,\;\;\;\;\;\;\;\;\;\;\; I=1,2,3\,,
\eea
with the parameters $\kappa_0$ breaking time reparameterizations, while $\lambda_0$ breaking space reparameterizations
 (as in \cite{Ricciardone:2016lym}).  The geometry is  de Sitter space, with metric ansatz given by \eqref{eq:desitter}.
   This system is a generalization of solid inflation \cite{Endlich:2012pz} to a supersolid set-up \cite{Bartolo:2015qvr, Ricciardone:2016lym}, where we break both time and space reparameterization invariance during inflation.
  The number of degrees of freedom generically changes:  since we break further symmetries,  a new scalar excitation is expected to propagate, 
  which we call  $\omega$, corresponding to the Goldstone
  boson for broken space reparameterization. 
   Such a quantity is associated with a perturbation of the field $\sigma^I$:
   \be
   \sigma^I\,=\,\lambda_0\,x^I+\frac{\lambda_0}{\sqrt{-\nabla^2}}\,\omega^I\,,
   \ee
   with $\omega^I\,=\,\hat \omega^I+\partial^I \omega$ (the former a vector, the latter the scalar fluctuation we are discussing). 
    On the other hand, our system is equipped with
  the scaling symmetry \eqref{eq:scalsymm}, that was shown in \cite{Domenech:2017kno} to forbid the propagation of  $\omega$ at large scales.
The argument, which we borrow from \cite{Domenech:2017kno}, can be summarized as follows:
 at  large scales, we can parameterize the configuration for $\omega^I$ as
 \be
 \label{eq:omega}
\omega^I\,=\,a \,\delta^I_{\,\,J}\,x^J+b^{IJ}\,\delta_{JM}\,x^M \,,
 \ee
 where $a$, $b^{IJ}$  are slowly varying in space, and $b$ is antisymmetric in its indexes. The first term proportional to $a$ in eq \eqref{eq:omega} is
 associated with a scalar
 mode, while the term proportional to $b^{IJ}$ corresponds to  a vector mode. However these fluctuations can be reabsorbed by an infinitesimal symmetry transformation
 that combines  rotational  \eqref{eq:rotsymm} with scale \eqref{eq:scalsymm} invariance. Hence, the mode $\omega^I$ does not acquires dynamics, at least  at large scales, being a pure gauge
 for the system under consideration.

The dynamics  of $\omega$ at smaller scales can be in principle taken into full account, as done in~\cite{Bartolo:2015qvr, Ricciardone:2016lym}, but gives only subleading contributions to the physics
 of the scalar sector  at large scales.  Since our work mostly focuses on the dynamics of tensor modes, for simplicity  we choose the available parameters such that $\omega$ does not propagate at all at quadratic level (i.e. its quadratic action has vanishing overall coefficient). This amounts to select parameters such that 
the following relation is satisfied 
\bea \label{condq2}
q_{\phi}\,q_{\sigma}&=&{\left( q_A+2 q_B\,\lambda_0^2\right)}\,,
\eea
and we assume in what follows that both $q_{\phi}$ and $q_{\sigma}$ are non-vanishing \footnote{ 
We also checked that  renouncing to this condition one finds a system where $\omega$ 
 acquires a healthy dynamics  free of Ostrogradsky instabilities  (for suitable choices of  parameters).}.
 
 We choose a branch among the possible background solutions,  where the value of the Hubble parameter is controlled by the parameters
 $q_i$, $(i=\phi,\sigma)$, generalizing the results of Section \ref{subsec:timebreak}. The background equations are solved by the following values of $H_0$,
 $\kappa_0$ (together with condition \eqref{condq2}):
\bea
H_0^2&=&\frac{M_{Pl}^2}{6\,q_{\sigma}}\,\left( q_A+2 q_B\,\lambda_0^2\right)\,,
\\
\kappa_0^4&=&\frac{M_{Pl}^4}{6}\,\lambda_0^2\, \left(q_C-9 q_B \lambda_0^2 \right)+2 M_{Pl}^2 v_1^2 \,.
\eea

Such a background configuration spontaneously breaks all space-time isometries, if the parameters $ \lambda_0$ and $\kappa_0$ are non-vanishing.  
The dynamics of fluctuations has interesting  features, above all in the tensor sector.  Using the same
 definition for tensor fluctuations as in eq \eqref{exp-gam}, we find the following quadratic action for tensor modes
\bea
S_{\gamma}^{(2)}\,=\,\int d t \,d^3  x\,a^3\,\frac{{ \cal N}_T}{2}
\left[ \dot \gamma_{ij}^2-c_T^2\,\left(\partial_k \gamma_{ij} \right)^2
-m_T^2 \gamma_{ij}^2
\right]\,,
\eea
where the parameters entering in the action are
\bea
 { \cal N}_T&=&
 \frac{M_{Pl}^2 \left(
 18 H_0^2+12 v_1^2+M_{Pl}^2\,\lambda_0^2\, \left( q_C-9\,\lambda_0^2\,q_B\right)
 \right)}{18\,H_0^2\,\left(3-c_T^2\right)}\,,\label{eq:tensorquant1}
 \\
 c_T^2&=&\frac{
 36 H_0^2-12 v_1^2-M_{Pl}^2\,\lambda_0^2\, \left( q_C+18 q_A +27 q_B \lambda_0^2 \right)
 }{
 36 H_0^2+12 v_1^2+M_{Pl}^2\,\,\lambda_0^2\,\, \left(q_C-6 q_A -21 q_B \lambda_0^2
 \right)}\,,\label{eq:citi}
 \\
 m_T^2&=&\frac{4\left(3-c_T^2\right)\,M_{Pl}^2\, H_0^2 \, \lambda_0^2\, \left(
 q_C
 +9 q_B \lambda_0^2 \right)}{18 H_0^2+12 v_1^2+M_{Pl}^2\,\lambda_0^2\, \left(q_C-9 q_B \lambda_0^2  \right)}\,.\label{eq:tensorquant3}
 \eea
 We recover the standard result setting $v_1\,=\,\lambda_0\,=\,0$. 
The spontaneous breaking of space isometries allows for a mass  term for  tensor fluctuations. It can be positive or negative, depending
on the sign and on the size of the parameters $q_C$, $q_B$, $v_1^2$.  Tensor fluctuations are not adiabatic modes in our 
scenario, hence they are not necessarily conserved at superhorizon scales:
if $m_T^2\,>\,0$, we find a {\it blue spectrum} ($n_T\,>\,0$) for primordial gravitational waves. We want to recall that also  the original set-up of solid inflation finds  a blue spectrum
for tensor modes: the spectral tilt in that case is however proportional to slow-roll parameters, hence its size is  small. In our scenario, 
we have more freedom to choose a parametrically larger value for the tensor parameters. 

\bigskip

The dynamics of scalar fluctuations is also relevant for our arguments. As usual, we need 
to satisfy ADM constraints~\cite{Maldacena:2002vr}, which we discuss in Appendix \ref{appendix}. In a convenient unitary gauge for the field $\phi$,   we find a single propagating scalar
mode, the field $A$ in eq \eqref{eq:metric}, and the corresponding quadratic action reads  
\bea
\label{eq:quadscal}
S_A^{(2)}\,=\,\frac{{ \cal N}_A}{2}\,\int d^4 x\,a^3\,
\left[ \dot A^2-c_A^2\,\left(\partial_i A \right)^2
-m_A^2 A^2
\right]\,.
\eea
The quantities ${ \cal N}_A, c_A, m_A$ are constant parameters given by
 \bea \label{defoNA}
 { \cal N}_A&=&
 12 \frac{\left(1-c_T^2\right)^2}{\left(2-c_T^2\right)^2}
\,  { \cal N}_T\,,
 \\
 c_A^2&=&\frac{2- c_T^2}{3}\,,
 \\
 m_A^2&=&m_T^2
 \,\frac{9 \left(2-c_T^2 \right)^2 q_B \lambda_0^2}{\left(1-c_T^2 \right)^2 \left(2 q_C +27 q_B \lambda_0^2 \right)}\,.
  \eea
 For make easier comparison, we expressed the  results using the tensor quantities ${ \cal N}_T$, $c_T$, $m_T$ defined
 in equations \eqref{eq:tensorquant1}, \eqref{eq:citi} and \eqref{eq:tensorquant3}.   As long as ${ \cal N}_T$ is positive, 
  the overall coefficient ${\cal N}_A$ of scalar fluctuations is positive (or at most vanishing when $c_T=1$): we can have a positive squared
  mass for the tensor modes in pure de Sitter space, without ghosts in the scalar sector, evading  the Higuchi bound. This is a relevant  
  feature of our scenario.
  We interpret this result as due to non-minimal interactions of scalar fields with gravity, which allow us to find a de Sitter background with non-vanishing
 background  $vevs$ for the scalar fields. Those $vevs$ spontaneously break all de Sitter symmetries, including the symmetries 
 crucial to prove the existence of the Higuchi ghost~\cite{Higuchi:1986py}, as pointed out in  \cite{Bordin:2016ruc}.
  Hence,  in our set-up with large couplings between tensor fluctuations 
  and the fields driving  accelerated  expansion, it
 is possible to consistently have  tensor fluctuations around pure de Sitter with masses in the interval $0<m_T^2<2 H_0^2$.

It would also be interesting to cosider higher order (e.g. cubic) interactions for scalar  modes, to estimate the strong coupling scale in this system, as well as 
 possible interesting features in scalar interactions. 

 \section{Phenomenological consequences}
 \label{sec:pheno}
 
 In this section we discuss the phenomenological consequences of our previous findings for what respects tensor modes, both at the level of power
 spectrum and bispectrum.
  We only focus on the tensor sector, although a system with broken space reparameterizations can have interesting consequences also
  for correlation functions involving the scalar sector (see e.g. \cite{Dimastrogiovanni:2014ina, Bartolo:2015qvr,Meerburg:2016ecv, Domenech:2017kno,
Endlich:2013jia,Akhshik:2014bla, Biagetti:2013kwa, Biagetti:2017viz} and \cite{solphen}).  We 
  are interested to consider  a set-up 
   where  the primordial tensor power spectrum increases towards small scales: it
   is unobservable at CMB scales, but it can be relevant at frequencies probed by interferometers ($10^{-4}$ Hz  $\lesssim f  \lesssim$ $10^3$ {\rm Hz}), where
   primordial scalar fluctuations are not important. To reduce the number of parameters, we  choose 
   $c_T=1$, $\lambda_0\,=\,1$,  and $q_B=0$ (while we continue to satisfy the condition \eqref{condq2}). 
  This choice  
    ``switches off'' completely the scalar sector at quadratic level (see eq \eqref{defoNA}), and  we are left
 with phenomenologically rich set up for tensor fluctuations around de Sitter space. The system still spontaneously breaks  time
 and space reparameterizations: given the 
 conditions we have analysed in the previous Section, and the fact that $\lambda_0=1$, we can take as symmetry breaking parameters 
 the quantities $\kappa_0$ and $m_T$, 
   or alternatively the
 parameters $q_A$, $q_C$ which appear in the initial action \eqref{eq:supaction}. 
 
We start discussing the consequences of our findings for the scale dependence of the tensor spectrum  that, as we have mentioned above, can have a positive tensor tilt
and can be detectable with interferometers as LISA~\cite{AmaroSeoane:2012km} (see also
\cite{earlydirect} for the first investigations on the possibility of direct detection of primordial gravitational waves).
Then we continue discussing how our system, which allows for large tensor non-Gaussianity enhanced in the squeezed limit, can induce a quadrupolar anisotropy in the tensor power spectrum. 
 
 \subsection{Tensor blue spectrum: inflationary tensor modes at interferometers}
 \label{subsec:spectrum}

When $c_T\,=\,1\,=\,\lambda_0$, and $q_B=0$, the quadratic action for 
 tensor modes  can be expressed as
\bea
S_{\gamma}^{(2)}\,=\,\frac{ \bar M^2_{Pl}}{4}\,\int d t d^3  x\,a^3\,
\left[ \dot \gamma_{ij}^2-\,\left(\partial_k \gamma_{ij} \right)^2
-m_T^2 \gamma_{ij}^2
\right]\,.
\eea
The effective Planck mass is given by 
\be
\label{eq:renplanck}
\bar M_{Pl}^2\,=\,M_{Pl}^2\,\left(1-
\frac{q_A\,M_{Pl}^2}{3 H_0^2}
\right)\,,
\ee
while the tensor mass squared  reads 
\be
m_T^2\,=\,\frac{4 H_0^2\,M_{Pl}^2\,q_C}{9 H_0^2 -3 q_A M_{Pl}^2}\,.
\ee
These quantities depend  on the two parameters $q_A$ and $q_C$ that are associated to the breaking of space reparameterization: when they are
set to zero, we recover the standard results for tensor fluctuations around a de Sitter background~\cite{Riotto:2002yw}.

We expand
tensor fluctuations in Fourier space  as
\be
\gamma_{ij}(t,\vec x)\,=\,\int\frac{d^3 k}{(2 \pi)^3}
\,\tilde{\gamma}_{ij}(t,\vec k)\,e^{i\,\vec k \cdot \vec x}\,.
\ee
The Fourier mode $\tilde \gamma_{ij}$ can be quantized and decomposed in terms of polarization tensors, and creation/annihilation operators
 \be
 \tilde \gamma_{ij}\,=\,\sum_s\,\left[\gamma_k\,{\bf e}_{ij}^{(s)}(\vec k)\,a_{s}(\vec k)+ \gamma_{-k}^*\,{\bf e}_{ij}^{*\,(s)}(-\vec k)\,a^\dagger_{s}(-\vec k)\right]\,,
 \ee
 with ${\bf e}_{ij}^{(s)}$ indicating the polarization tensor with helicity $s=\pm2$, satisfying the transverse-traceless condition $\,k_i \,{\bf e}_{ij}^{(s)}\,=\,{\bf e}_{ii}^{(s)}=\,0$.  
 We adopt the normalization conditions: ${\bf e}_{ij}^{(s)}\,{\bf e}_{ij}^{(s')}\,=\,\delta_{s s'}$. We also use the following property ${\bf e}_{ij}^{*\,(s)}(\vec k)\,=\,{\bf e}_{ij}^{(-s)}(\vec k)\,=\,{\bf e}_{ij}^{(s)}(-\vec k)$. The creation/annihilation operators satisfy the usual commutation relations $[a_{s}(\vec k),a^\dagger_{s'}(\vec k')]=(2\pi)^3\delta_{ss'}\delta^{(3)}(\vec{k}-\vec{k}')$.

 \begin{figure}[t!]
  \begin{center}
\includegraphics[width=6.4cm]{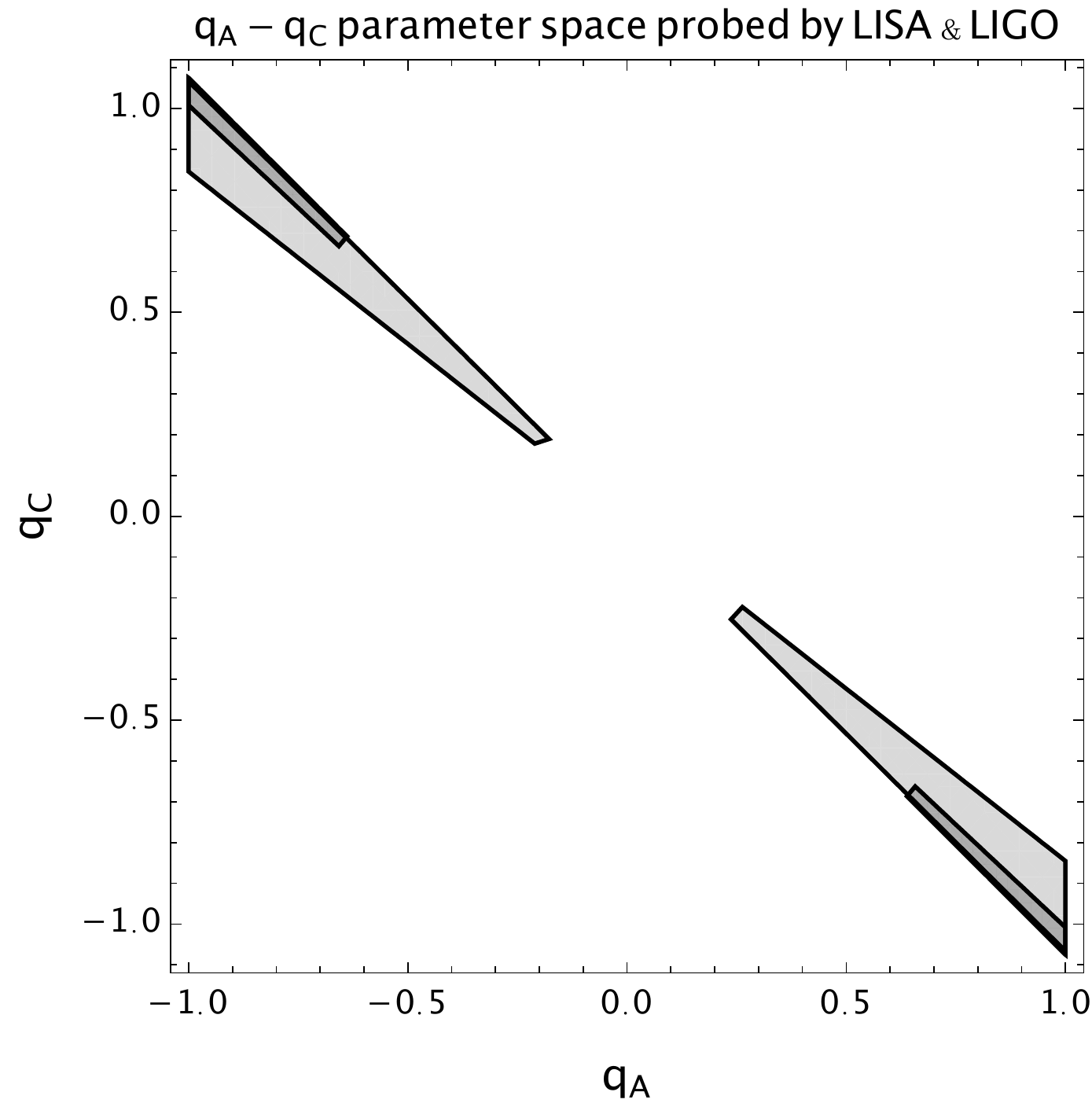}
\includegraphics[width=9.85cm]{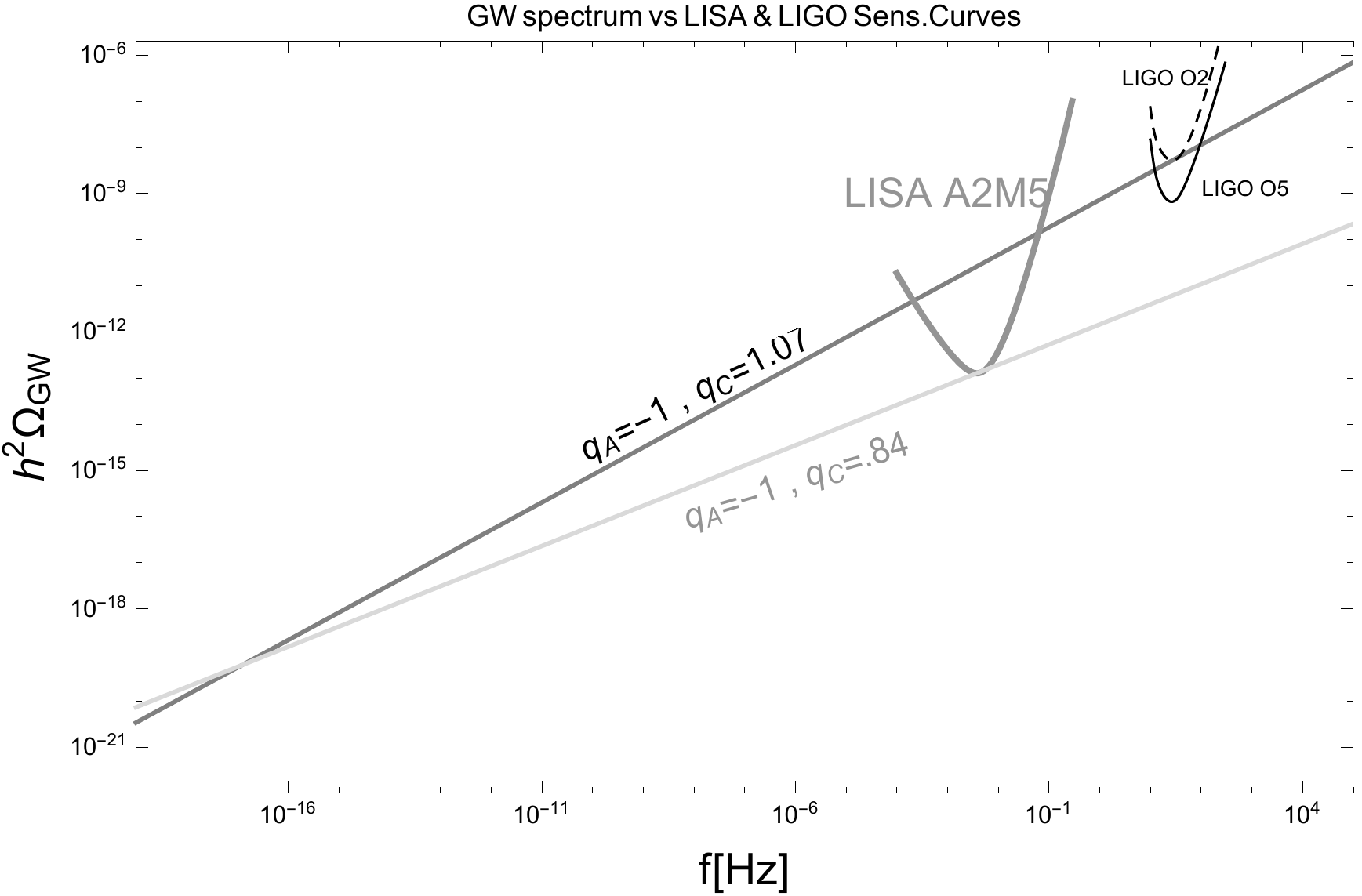}
\end{center}
\caption{{\bf Left}: Parameter space probed by LISA and LIGO for $q_{A}$-$q_{C}$ parameters. The lighter gray region is the region probed by LISA in the A$2$M$5$ (5 years mission, 2 million armlength). While the darker gray is the region probed by LIGO considering a detection in the O5 run. The plot is obtained for $H=10^{12}$ GeV. {\bf Right:} Spectrum of GWs energy density $h^{2}\Omega_{gw}$ for different values of $q_{A}$ and $q_{C}$ parameters, compared with the sensitivity of LISA  A$2$M$5$ (gray curves) and Advanced LIGO (black curves) detectors. We use $k_{*}=0.002~{\rm{Mpc^{-1}}}$ as a pivot scale.}
     \label{fig: superGWLISA}
\end{figure}

Quantization proceeds in a textbook manner \cite{Lyth:2009zz}, and one finds  
the following expression for 
the tensor power spectrum ${\cal P}_\gamma\,=\,(k^3/4 \pi^2)\,\langle \gamma^{s}  \gamma^{s} \rangle$ 
 (the two-point (2-pt) function has the same amplitude for each of the two polarizations $+$ and $\times$)
\be
\label{eq:tensps}
{\cal P}_\gamma\,=\,\frac{H_0^2}{\bar M_{Pl}^2}\,\frac{2^{2 \nu-2}}{\pi^2}
\,\left(\frac{k}{k_*}\right)^{3-2\nu}\,,
\ee
where  $\bar{M}_{Pl}$ is the renormalized Planck mass~\eqref{eq:renplanck}, the tensor spectral tilt $n_{T} \equiv \frac{d\,\ln{ {\cal P}_h}}{d\,\ln{k}}\,\Big|_{ k\,=\,k_*}=3-2\nu$ and
\be
\label{eq:nu}
\nu\,=\,\sqrt{\frac94-\frac{m_T^2}{H_0^2}}\,.
\ee

We then find that in our set-up, when $m_T^2>0$ and hence $(3-2\nu)\,>\,0$, we can have a power spectrum for tensor modes which increases towards small scales (a blue spectrum), where interferometers like LISA~\cite{AmaroSeoane:2012km} and LIGO-Virgo are sensititve~\cite{Harry:2010zz}.
In Fig.~\ref{fig: superGWLISA} we represent the capability of experiments like LISA and Advanced LIGO (run O2 and future O5) to probe the parameter space of the model under analysis. In particular, on the left plot, we show how a LISA configuration with 5 years mission and 2 million km arm-length (which is the one closely similar to the ESA approved one), can probe 
 the parameters of our specific inflationary model. We can also appreciate how LISA is sensitive to a larger range of parameters compared to LIGO.
In the right plot we have plotted the GW energy density for some representative values of the parameters $q_{A}$ and $q_{C}$ versus the LISA and Advanced LIGO sensitivities. The fractional GW energy density is related to the power spectrum \eqref{eq:tensps} by the transfer function (see e.g.~\cite{Kuroyanagi:2014nba, Maggiore:1999vm} for more details). Our scenario allows for an increasing GW power on small scales,
 and its amplitude results well within LISA sensitivity curves. 

 \subsection{Tensor squeezed non-Gaussianity and anisotropic  tensor power spectrum}
\label{subsec:ng}

Since tensor modes are non-adiabatic in this system, we expect important consequences
for tensor interactions and, in particular, for primordial tensor non-Gaussianity, including a possible violation of Maldacena's
consistency condition~\cite{Maldacena:2011nz}.  We discuss this subject in this section,  analysing   implications 
for the tensor power spectrum.

The action for tensor fluctuations expanded to cubic order is (we choose as above $c_T\,=\,1$, $q_B=0$)
\bea
S_{\gamma}^{(3)}&=&\frac{
\bar M_{Pl}^2}{4}\int d t d^3 x\,a^3\,\left[\frac{{\cal C}_1}{a^2}
\,\gamma_{ij} \gamma_{nm}\left(\partial_j \partial_n \gamma_{im}-\frac12 \partial_i \partial_j \gamma_{mn} \right)
+
{\cal C}_2\,\gamma_{ij} \gamma_{jm} \gamma_{mi}
 \right] \,,\label{accubt}
\eea
where the two coefficients ${\cal C}_1$ and ${\cal C}_2$ are equal to
\bea
{\cal C}_1&=&
1+\frac{M_{Pl}^2}{36\,H_0^2}
\left(q_C-30 q_A\right)\,,
\\    
{\cal C}_2&=&\,
\frac{2\,M_{Pl}^2}{27}
\left( 4 q_C+9 q_A\right)\,.
\eea
Action \eqref{accubt} is weighted by the renormalized Planck mass $\bar M_{Pl}^2$ of eq \eqref{eq:renplanck}. 
The first part of action  \eqref{accubt}, weighted by the parameter ${\cal C}_1$, has the very same structure as the third order action of tensor modes
in General Relativity (see e.g. \cite{Maldacena:2011nz,Gao:2011vs}), but the overall coefficient is different. In the limit where we restore space  reparameterization
invariance, setting $q_A\,=\,q_C\,=\,0$,  we
find ${\cal C}_1\,=\,1$, which is  the General Relativity result. However in general, given our freedom to choose the parameter  
$q_A$, $q_C$
we can have parametrically large deviations from  ${\cal C}_1\,=\,1$. The second contribution to 
action  \eqref{accubt}, weighted by ${\cal C}_2$, is absent in GR and in any single field inflation model, and
is characteristic of theories which break space diffeomorphisms: when $q_A\,=\,q_C\,=\,0$,   we
find ${\cal C}_2\,=\,0$.

\begin{figure}[t!]
  \begin{center}
    \includegraphics[keepaspectratio=true,height=60mm]{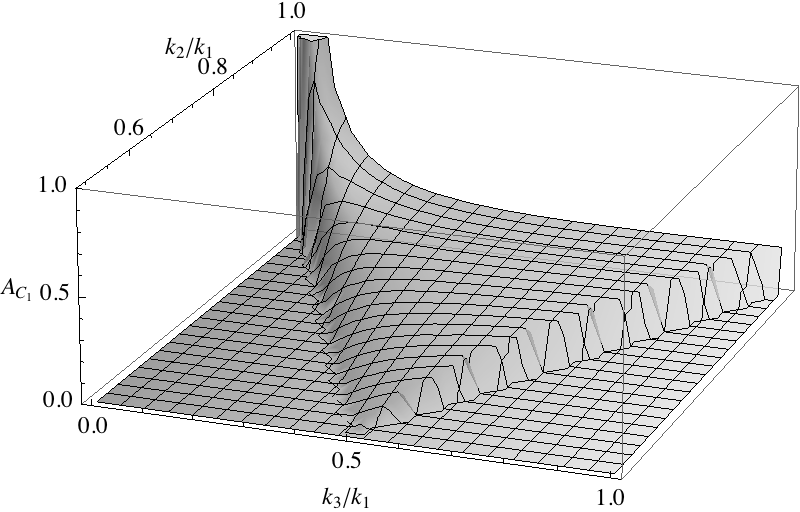}
  \end{center}
 \caption{Plot of the shape function of the tensor bispectrum for the term proportional to ${\cal C}_1$: ${\cal A}_{{\cal C}_{1}}^{+++}(1, k_2/k_1, k_3/k_1)(k_2/k_1)^{2}(k_3/k_1)^{2}$
  as a function of $k_2/k_1$ and $k_3/k_1$. The plot is normalized to unity for
  equilateral configurations $k_2/k_1=k_3/k_1=1$. }
  \label{fig:C1shape}
\end{figure}

A distinctive feature of the cubic operators contained in action \eqref{accubt} is that they lead to tensor
non-Gaussianity (nG) with shape peaked in the squeezed limit (in contrast with the ``equilateral shape''
tensor nG characteristic of systems with particle production, see e.g.~\cite{Barnaby:2011vw,Barnaby:2010vf,Cook:2013xea}, or models
 of single field inflation~\cite{Maldacena:2011nz,Soda:2011am,Gao:2011vs}). This subject has been developed
in our previous paper~\cite{Ricciardone:2016lym} to which we refer the reader for more details:  here we only present in
Fig.~\ref{fig:C1shape} a plot of the shape function 
of the bispectrum  (Fourier transform of the 3-pt correlation function) for tensor modes, associated with the operator proportional to ${\cal C}_1$,
whose shape is manifestly peaked in the squeezed limit. We reiterate that its amplitude can 
 be parametrically larger than in GR, by choosing appropriately the parameters $q_A$, $q_C$.

\smallskip
Tensor non-Gaussianity might appear as a futuristic observable well beyond the
 sensitivity of planned experiments. However
   a large non-Gaussian amplitude peaked in the squeezed limit can induce a large quadrupolar anisotropy in the tensor power
   spectrum, leading to a possible  estimator for tensor non-Gaussianity. 
Indeed, we can write the following general expression for 
 the squeezed limit of the tensor 3-pt function

\be
\label{eq:squeezedlimit}
\lim_{\vec q\to 0}\,
\langle \gamma_{\vec{q}}^{s_1} \,\gamma_{\vec{k}}^{s_2} \,\gamma_{-\vec{k}}^{s_3} \,\rangle'
\,=\,\delta^{s_2\,s_3}\,{\cal P}_\gamma(q)\,{\cal P}_\gamma(k)\,\left(\frac32+f_{\rm NL}^T\right)\,
{\bf e}_{ij}^{(s_1)}(\vec q)\,\frac{k^i\,k^j}{k^2}\,.
\ee
where $q$ and $k$ represent the momenta of the {\it long} and {\it short} tensor mode, respectively. The non-linear parameter $f_{\rm NL}^T$
 characterises how much we 
 are  departing  from the standard consistency relation, due to non-adiabaticity of tensor modes. For example in our case, setting ${\cal C}_2=0$, 
   $f_{\rm NL}^T$ is proportional to $\left({\cal C}_1-1\right)$, which can be parametrically larger than unity. 
 In the presence of large squeezed tensor non-Gaussian signal, 
a single long wavelength tensor mode -- which we denote with a bar as
$\bar \gamma_{\vec{q}}^{s}$ --
 modulates
the tensor two-point function as follows~\cite{tenslongA,Bordin:2016ruc}
 \be
 \langle \gamma_{\vec{k}}^{s_1} \,\gamma_{-
 \vec{k}}^{s_2}  \rangle'_{\bar \gamma}\,=\,\langle \gamma_{\vec{k}}^{s_1} \,\gamma_{-\vec{k}}^{s_2}  \rangle'_0
 +\bar \gamma_{\vec{q}}^{s_3}\,\frac{\langle \bar \gamma_{\vec{q}}^{s_3} \gamma_{\vec{k}}^{s_1} \,\gamma_{-\vec{k}}^{s_2} \rangle'}{{\cal P}_\gamma(q)}\,,
 \hskip1cm{\text{for}} \hskip1cm q\ll k\,.
 \ee
 We denote with $\langle\dots  \rangle'_{\bar \gamma}$ the n-pt function modulated
 by the long tensor mode, while  with $\langle\dots  \rangle'_{0}$ the un-modulated quantity (the prime means that we do not include the momentum conserving
 $\delta$-function in these expressions.).

 Physically, we should
 consider the cumulative effect
 of  {\it all} soft graviton modes
  whose  momenta  are smaller than a representative momentum $q_L$, which is  proportional to the inverse of the size
 of
 the instrument
  we use to detect tensor modes; in the extreme case, the size of the apparatus can be the entire observable universe, of order  $H^{-1}$.
  At a given position $\vec x$, we can write (see e.g.~\cite{inNGa})
 \be
  \langle \gamma_{\vec{k}_1}^{s_1} \,\gamma_{\vec{k}_2}^{s_2}  \rangle'_{\bar \gamma} (\vec x)\,=\,\langle \gamma_{\vec{k}_1}^{s_1} \,\gamma_{\vec{k}_2}^{s_2}  \rangle'_0
 +\frac{1}{V_L}
 \int_{|\vec q|<q_L}\,d^3 q\,e^{i \vec q \cdot \vec x }
 \left(\sum_{s_3\,=\,+,\times}
 \bar \gamma_{\vec{q}}^{s_3}\,\frac{\langle \bar \gamma_{\vec{q}}^{s_3} \gamma_{\vec{k}_1}^{s_1} \,\gamma_{\vec{k}_2}^{s_2} \rangle'}{{\cal P}_\gamma(q)}
\right)\,,
 \ee
 with $V_L$ the volume (in Fourier space) of the space of 3-momenta over which we are integrating. 
We can substitute above the expression for the squeezed limit of the 3-pt function, eq \eqref{eq:squeezedlimit}, and we get
\bea
  \langle \gamma_{\vec{k}}^{s_1} \,\gamma_{-\vec{k}}^{s_2}  \rangle'_{\bar \gamma} (\vec x)&=&\langle \gamma_{\vec{k}}^{s_1} \,\gamma_{-\vec{k}}^{s_2}  \rangle'_0\left(1+
  f_{NL}^T\,\frac{k^i\,k^j}{k^2}\,\frac{1}{V_L}
\,\int_{|\vec q|<q_L}\,d^3 q\,e^{i \vec q \cdot \vec x }
  \sum_{s_3}{\bf e}_{ij}^{(s_3)}(\vec q)   \bar \gamma_q^{(s_3)}
\right)\,,
\\
&=& \langle \gamma_{\vec{k}}^{s_1} \,\gamma_{-\vec{k}}^{s_2}  \rangle'_0\,\cdot\,\left(1+
\frac{k^i\,k^j}{k^2}\,{\cal Q}_{ij}(\vec x)
\right)\,, \label{quadasym}
 \eea
 where we have introduced the quantity 
 \be
 \label{eq:quij}
 {\cal Q}_{ij}(\vec x)\,=\,\frac{ f_{NL}^T}{V_L}
\,\int_{|\vec q|<q_L}\,d^3 q\,e^{i \vec q \cdot \vec x }
 \sum_s  {\bf e}_{ij}^{(s)}(\vec q) \bar   \gamma_q^{(s)}\,.
 \ee
 
 Hence eq \eqref{quadasym} implies that the tensor power spectrum 
  can acquire a position-dependent quadrupolar asymmetry, due the coupling between  long and short wavelength tensor
  fluctuations in the primordial universe. 
 The  quantity ${\cal Q}_{ij}$ acquires stochastic random values over a Gaussian distribution, with variance
 that depends on the nG parameter appearing in eq \eqref{eq:squeezedlimit} (see e.g. \cite{Bordin:2016ruc}):
\be \label{Qvariance}
\langle  {\cal Q}_{ij}  {\cal Q}_{ij} \rangle\,=\,\frac{4 \pi}{V_L^2}\,\left(f_{NL}^T \right)^2\,\sum_s\,\int_{q<q_L}
\, d q\,q^2\, \langle   \bar\gamma_{\vec q}^{(s)}\bar \gamma_{-\vec q}^{(s)} \rangle'\,.
\ee
As discussed in the previous subsection, 
we are interested in a situation where  the tensor tilt 
 $n_T$ is
  positive, 
   implying that the integral in eq \eqref{Qvariance} 
 converges in the infrared domain. Notice that  although the tensor 2pt function acquires a quadrupolar anisotropy, still this effect 
 is the same for both polarizations $+$ and $\times$, and we do not have any cross correlation among polarizations: this is due to the fact
 that the background is still isotropic (being de Sitter space), and the effect described above is only due to
 the  coupling between 
 long and short modes, which is induced by non-Gaussianity. As a consequence,
 the tensor power spectrum can be parameterized as
 \be \label{anipsf}
 {\cal P}_{\bar \gamma} (k)\,=\,{\cal P}_0 (k)\,\left(1+{\cal Q}_{ij}\,\frac{k^i\,k^j}{k^2} \right)\,.
 \ee
 where ${\cal P}_0 (k)$ represents the standard isotropic contribution to the tensor power spectrum. 
In order to  characterize the SGWB, since GWs have two polarizations, is common to expand them in terms of Stokes parameters, by analogy with electromagnetic cases; these make more clear the properties and symmetries of GW polarizations. In particular, in our specific case,  when the tensor  two point function is decomposed in terms of Stokes parameters the only non-vanishing
 parameter appears to be the intensity $I$ of GWs, characterised by  a quadrupolar anisotropy controlled by the quantity ${\cal Q}_{ij}$.
 The other Stokes parameters $V, U$ and $Q$ are zero: the first because in this case we do not have any asymmetry in the 
 two polarization amplitudes of the GWs, typical of models that violate parity symmetry~\cite{Shiraishi:2013kxa}; $Q$ and $U$ because 
 they carry additional information about linear polarizations, in particular about cross correlations
  among polarizations which are absent
   in our specific case.   
  
\subsection{Interferometer response to an  anisotropic tensor power spectrum}
\label{subsec:interferometer}

 For the rest of this section, we briefly start to explore consequences of the results found so far,  for
what respect the power spectrum  of primordial tensor 
 modes detectable  with gravitational wave interferometers. 
 We use the methods first discussed in \cite{Cornish:2001bb}, and developed
in various works as~\cite{Kato:2015bye,Smith:2016jqs, Thorne:2017jft}. We follow closely~\cite{Smith:2016jqs}, extending their results to describe
our set-up,  which includes an anisotropic tensor power spectrum. 

We focus our attention only on the consequences of 
 tensor anisotropies of the form of eq \eqref{anipsf} for the relative phase shift of light which travels between test masses
located at the extremes of  the arms of an interferometer.  We postpone a more detailed analysis on the physical implications and characterisation of 
 the GW signal to  future publications. \\
     
  It is convenient to expand gravitational wave modes in plane waves, implementing an interferometer  notation (see e.g. \cite{Allen:1996vm})
\be
\gamma_{ij}(t,\vec x)\,=\,\int_{-\infty}^{+\infty}\,d f\,\int d^2 \vec n\,\sum_{s}\, \gamma^{(s)}(f,\,\vec n)\,{\bf e}_{ij}^{(s)}\,e^{i\,2 \pi\,f\left(t-\vec n \cdot \vec x\right)}\,,
\ee
with $f$ the frequency, and $\vec n$ the versor in the direction of 3-momentum of the propagating wave. 
In our specific set-up,  the 2-pt function for the mode function appearing the previous expansion reads
 \be
 \langle
  \gamma^{(s)}(f,\,\vec n),\,  \gamma^{(s')}(f',\,\vec n')
  \rangle
  \,=\,\frac{\delta(f-f')}{2}\, \,\frac{\delta^{(2)} (\vec n+\vec n')}{4 \pi}
\,\delta^{s\,s'} S_{\gamma}(f)\,\left( 1+{\cal Q}_{mn} n^m n^n \right)\,, \label{2ptgin}
   \ee
   where $ S_{\gamma}(f)$ is the  amplitude of the intensity signal, which depends on the amplitude of the primordial tensor spectrum
   \be
  S_{\gamma}(f)\,=\,\frac{H_0^2}{ {\bar M}_{Pl}^2\,f^3}\,\frac{2^{2 \nu-2}}{\pi^2} \left( \frac{f}{f_*}\right)^{3-2\nu}\,,
   \ee
   with $\nu$ defined in eq \eqref{eq:nu}, 
  while 
 ${\cal Q}_{mn}$ is the anisotropic contribution to the tensor power spectrum defined in eq \eqref{eq:quij}. 
 For simplicity,  in this work  we make the hypothesis that the quantity ${\cal Q}_{mn}$ is constant, and independent
from  the position of the interferometer. This situation occurs for example if the spatial dependence of the long wavelength tensor mode controlling ${\cal Q}_{mn}$ is  weak
 within the horizon corresponding to region of space
 containing the instrument (for example the solar system in the case of LISA).

We model  an interferometer as made of $n$ arms, with test masses $M_{i}$ at their extremes ($i=1,\dots 2n$), located
at position $\vec x_i$. 
The basic quantity which controls how the interferometer responds to a gravitational wave is 
 the electromagnetic phase shift  accumulated by light during its travelling  along an arm of 
 the instrument:
 \be
 \phi_{12}(t)\,=\,\phi_0 \left[ 
 1+ 
 \int_{-\infty}^{+\infty}\,d f\,\int d^2 \vec n\,\sum_{s}\, \gamma^{(s)}\,{\bf e}_{ab}^{(s)}\,e^{i\,2 \pi\,f\left(t-\vec n \cdot \vec x_1\right)}
\,{\cal D}_{ab}\left(f,\,\vec \ell_{12} \cdot \vec n \right)
 \right]\,,
 \ee
 with $\vec x_1$ the location of mass $M_1$, and $\vec x_2\,=\,\vec x_1+
 L\,\vec \ell_{12}$ the location of mass $M_2$ ($\vec \ell_{12}$ being the unit vector in the direction of the interferometer
 arm). $\phi_0$ is the phase measured in absence of a gravitational wave passing through the arms of the interferometer.  The quantity ${\cal D}_{ab}\left(f,\,\vec \ell_{12} \cdot \vec n \right)$ is the arm transfer function
 \be
 {\cal D}_{ab}\left(f,\,\vec \ell \cdot \vec n \right)
 \,=\,\frac12\,
  \ell_a\, \ell_b\,
{\cal M} \left(f,\,\vec \ell \cdot \vec n \right)\,,
 \ee 
 with ${\cal M} $  given by 
  $$
  {\cal M}(\vec \ell \cdot \vec n, \,f)\,\equiv\,\frac{i}{2\pi\,L_D\,f}\frac{\exp{\left[
  2 \pi\,i\,L_D\,f\left(1-\vec \ell\cdot \vec n\right)\right]} -1 }{1-\vec \ell \cdot \vec n}\,,
  $$
 and $1/(2 \pi L_D)$ is the characteristic frequency scale of the detector.  The signal  $s_1(t)$,  associated with the interferometer arm, is defined in terms of the phase shift of the electromagnetic wave as it travels from one end of the arm to the other, and back \cite{Thorne:2017jft}
  \be
  s_1(t)\,=\,\phi_{12}(t-2 L)+\phi_{21} (t-L)+n_1(t)\,,
  \ee
  where with $\phi_{12}$ and $\phi_{21}$ we indicate the shifts and  with $n_1(t)$ a noise term. Hence we learn that the signal 
 is built in terms of a linear combination of phase shifts: it is essential to specify the properties of the latter, in order to deduce implications for the former.\\ 
 It is convenient to work with 
  the Fourier transform of the phase shift accumulated along any of the interferometer arms:
  \be
\Delta \tilde \phi_{ij}(f)\,=\,
 \int d^2 \vec n\,\sum_{s}\, \gamma^{(s)}\,{\bf e}_{ab}^{(s)}\,e^{-i\,2 \pi\,f \vec n \cdot \vec x_1}
\,{\cal D}_{ab}\left(f,\,\vec \ell_{ij} \cdot \vec n \right)\,.
 \ee
 To extract information about the stochastic GW background we need to correlate two phase shifts, so, using eq \eqref{2ptgin}, we find that the 2-pt function of the fase shift reads as
\be
\langle \Delta \tilde \phi_{ij}(f)  \Delta \tilde \phi_{kl}^*(f')
\rangle\,=\, \frac12 \,\delta(f-f')
\,\delta^{s s'}\,{S}_{h}  (f) \,{\cal R}_{s\,s'}^{ij,\,kl} (f)\,,
\ee
 where the  quantity
  ${\cal R}_{s\,s'}^{ij,\,kl} (f)$ is the
   response function of the interferometer to a gravitational wave passing through its arms.  In our specific case it 
acquires a new contribution with respect to the standard case where just an isotropic stochastic signal is present~\cite{Romano:2016dpx, Smith:2016jqs}, which we collect
 in the second line of the following expression
\bea
{\cal R}_{s\,s'}^{ij,\,kl} (f)&\equiv&\int \frac{d^2 \vec n }{4 \pi}
\,e^{i\,2 \pi\,f\,\vec n \left( \vec x_i-\vec x_k\right)}\,{\cal D}_{ab}(\vec \ell_{ij} \cdot \vec n,\,f)
{\bf e}_{ab}^{(p)}(\vec n)\, 
{\cal D}^*_{cd}(\vec \ell_{kl} \cdot \vec n,\,f)
{\bf e}_{cd}^{(p')}(\vec n')
\nonumber
\\
&+&{\cal Q}_{m n} 
\int \frac{d^2 \vec n }{4 \pi} \,n_m n_{n}\, 
\,e^{i\,2 \pi\,f\,\vec n \left( \vec x_i-\vec x_k\right)}\,{\cal D}_{ab}(\vec \ell_{ij} \cdot \vec n,\,f)
{\bf e}_{ab}^{(p)}(\vec n)\, 
{\cal D}^*_{cd}(\vec \ell_{kl}\cdot \vec n,\,f)
{\bf e}_{cd}^{(p')}(\vec n)\,.
\label{Rfunction1}
\eea
The contribution proportional to the tensor  ${\cal Q}_{mn}$, in the second line of the previous formula,
  quantifies how the quadrupolar anisotropy affects the interferometer response. In this way the response function is given by the sum of two contributions  
\bea 
{\cal R}_{s\,s'}^{ij,\,kl} (f)
&\equiv&
^{(0)}\,{\cal R}_{s\,s'}^{ij,\,kl} (f)+
\,^{(1)}\,{\cal R}_{s\,s'}^{ij,\,kl,\,mn} (f)
\,\, {\cal Q}_{mn}\,.
\eea
    The first part, $^{(0)}\,{\cal R}$,  is the standard contribution discussed in \cite{Smith:2016jqs}. The second part, proportional  $^{(1)}\,{\cal R}$, characterises
  a new contribution 
   modulated by the  squeezed non-Gaussianity.  The integrals in 
   eq \eqref{Rfunction1}
   depend on the positions $\vec x_i$ and $\vec x_k$ of the masses $M_i$ and $M_k$ of the interferometer, as
   well as on  the orientation of the interferometer with respect to the ``preferred directions'' controlled by  
  the tensorial quantity ${\cal Q}_{mn} $. This latter  quantity   depends on the amplitude of  long-wavelength graviton modes whose wavelength is  
     larger than the size of the instrument used to make the measurement. For the case of LISA interferometer configuration, for example,  such size is the solar system.
      
  \smallskip
Collecting these results, we find that  the two-point function among the phase shifts  reads
\bea
\langle \Delta \tilde \phi_{ij}(f)  \Delta \tilde \phi_{kl}^*(f)
\rangle
&=&
 \frac{\delta^{s \,s'}\,S_{h}  (f) }{2}
\,^{(0)}{\cal R}_{s\,s'}^{ij,\,kl} (f)
+ \frac{\delta^{s \,s'} \,S_{h}  (f)}{2}\,
{\cal Q}_{mn}\, 
 \,^{(1)}{\cal R}_{s\,s'}^{ij,\,kl,\,mn} (f)\,.
\label{2pfaf}
\eea
The second term contains the anisotropic  contribution associated with tensor nG, and it is a distinctive feature
of our system. The value of this contribution depends 
on how the interferometer arm vectors $\vec \ell_{ij}$ are oriented with respect to the long wavelength tensor mode, through the quantity
${\cal Q}_{mn}$. Interestingly, while the first term in   \eqref{2pfaf} is constant in time (as long as the relative orientation
of the interferometer arms do not change) 
the value of the second contribution in  \eqref{2pfaf} can depend on time, since the orientation
of the interferometer arms  changes  with respect to ${\cal Q}_{mn}$ as time passes:
 few hours  for earth-based  interferometers (e.g. LIGO-Virgo), or few months  for space interferometers, like LISA.

 This fact can
 lead to distinctive   observational consequences. The two-point phase shift correlation is the basic ingredient for 
 building signal estimators for gravitational wave detection at interferometers~\cite{Cornish:2001bb,Kato:2015bye,Smith:2016jqs, Thorne:2017jft},
  in terms of correlation functions of the signal. 
 A measurement of
 a  time-dependent modulation 
   for a stochastic primordial tensor spectrum,
 whose amplitude changes with time (within months for an interferometer like LISA),  can  be a smoking gun
  for large non-Gaussianity in the tensor sector. 
   On the contrary, bounds on time variations of the amplitude of the stochastic signal 
   can be used to impose bounds on the size of non-Gaussianity parameters, like $f_{NL}^{T}$.    
  We plan to return to discuss in detail phenomenological consequences
  of these results in a separate
   publication.

\section{Discussion}
\label{sec:disc}

In this paper we have developed  a system of supersolid inflation, where time and space reparameterizations are spontaneously
 broken by the background $vevs$ of a set of 
scalar fields coupled to gravity. We shown that this scenario has distinctive features for the properties of primordial tensor modes. The primordial
tensor spectrum can have a blue tilt with parametrically large value of the tensor tilt $n_T$.
 The Higuchi bound can be avoided thanks to  interactions between the tensor perturbations and the fields driving inflation. Tensor modes are not adiabatic in this 
 set-up.  
 This system can have large tensor
non-Gaussianities enhanced in the squeezed limit, which couple long and short tensor fluctuations, leading to a quadrupolar anisotropy
in the tensor power spectrum. We have discussed phenomenological consequences of our findings, in particular  their
 implications for interferometer searches of a primordial stochastic gravitational wave background. 
 We shown that a future space-based interferometer mission like LISA can probe (and constrain)
 parameters of our model. We have briefly discussed how a quadrupolar anisotropy in the tensor
 power spectrum can lead to a time dependent modulation of  the amplitude of a
  SGWB. This fact can in principle allow one to build  a
  specific  estimator for testing squeezed tensor non-Gaussianities, to be used with future interferometers.

\subsection*{Acknowledgments}
It is a pleasure to  thank Marco Peloso for useful discussions. We also 
  would like to acknowledge the LISA Cosmology working group
  for discussions, and for feedback on a presentation of these results 
   during the IV LISA workshop in Mainz.
The work of GT is partially supported by the STFC grant ST/P00055X/1.

\newpage
\begin{appendix}
\section{ADM constraints for  scalar perturbations}
\label{appendix}
In this Appendix 
we discuss  the constraint conditions for the scalar sector of the supersolid inflation system described in 
Section \ref{subsec:timespacebreak}.  
The background metric and scalar fields are
\bea
d s^2&=&-d t^2+e^{2 H_0 t}\,d \vec x^2\,,
\\
\phi&=& \kappa_0^2\,t\,,
\\
\sigma^I&=& \lambda_0\,x^I\,.
\eea
We choose the parameter $q_\phi$ as 

\bea \label{condq2a}
q_\phi&=&\frac{\left( q_A+2 q_B\,\lambda_0^2\right)}{q_\sigma}\,.
\eea
The branch of 
background solutions we are interested in  determines 
the following choice for the available parameters

\bea
H_0^2&=&\frac{M_{Pl}^2}{6\,q_\sigma}\,\left( q_A+2 q_B\,\lambda_0^2\right)\,,
\\
\kappa^2_0&=&\frac{M_{Pl}^4}{6} \left(q_C-9 q_B \lambda_0^4 \right)+2 M_{Pl}^2 v_1^2 \,,
\eea

In order to study scalar fluctuations, we
  implement a partial unitary gauge, the same as in our work~\cite{Bartolo:2015qvr}. We leave the scalar $\phi$
unperturbed, while we perturb metric and scalars $\sigma^I$ as
\bea
d s^2&=&
-\left(1+2 N\right)\,
 d t^2+2\,e^{2\, H_0\,t}\,\partial_i B\,d x^i dt+e^{2\, H_0\,t}\,\left(1+2 A \right)\,\delta_{ij}\,d x^i d x^j\,,
\\
\sigma^I
&=&\lambda_0\,x^I+\frac{\lambda_0}{\sqrt{-\nabla^2}}\,\partial^I\,\omega\,.
\eea
The equations of motion for $N$, $B$, $\omega$ correspond to  constraint equations, which lead to the 
 following solutions in momentum space
\bea
N&=&\frac{\dot A}{H_0\left(2-c_T^2\right)}\,,
\\
B&=&-\frac{\left[3 \left(1-c_T^2\right)^2\,a^2\,H_0\,\dot{A}+\left(2-c_T^2 \right) 
\,k^2\,A
\right]}{\left(2-c_T^2\right)^2\,H_0\,a^2\,k^2}\,,
\\
\omega&=&-\left(
\frac{27\,q_B\,\lambda_0^4}{k\,\left( 2 q_C+27 q_B\,\lambda_0^4\right)}
\right)\,A\,,
\eea
with $c_T$ given in eq \eqref{eq:citi}. Substituting these expressions in the action, we find that the scalar
 dynamics corresponds to a single field scenario, controlled by the quadratic action \eqref{eq:quadscal}. 

\end{appendix}
\newpage



\begin{thebibliography}{99}



\bibitem{Maggiore:1999vm}
  M.~Maggiore,
  Phys.\ Rept.\  {\bf 331} (2000) 283
  [gr-qc/9909001].

\bibitem{Guzzetti:2016mkm}
  M.~Guzzetti, C., N.~Bartolo, Liguori, M. and S.~Matarrese,
  Riv.\ Nuovo Cim.\  {\bf 39} (2016) no.9,  399
  [arXiv:1605.01615 [astro-ph.CO]].



\bibitem{Kamionkowski:2015yta}
  M.~Kamionkowski and E.~D.~Kovetz,
  Ann.\ Rev.\ Astron.\ Astrophys.\  {\bf 54} (2016) 227
  [arXiv:1510.06042 [astro-ph.CO]].



\bibitem{Cook:2011hg}
  J.~L.~Cook and L.~Sorbo,
  Phys.\ Rev.\ D {\bf 85} (2012) 023534
  [arXiv:1109.0022 [astro-ph.CO]].
  
\bibitem{Cook:2013xea}
  J.~L.~Cook and L.~Sorbo,
  JCAP {\bf 1311} (2013) 047
  [arXiv:1307.7077 [astro-ph.CO]].


  
\bibitem{Biagetti:2014asa}
  M.~Biagetti, E.~Dimastrogiovanni, M.~Fasiello and M.~Peloso,
  JCAP {\bf 1504} (2015) 011
  [arXiv:1411.3029 [astro-ph.CO]].
  
\bibitem{Barnaby:2011qe}
  N.~Barnaby, E.~Pajer and M.~Peloso,
  Phys.\ Rev.\ D {\bf 85} (2012) 023525
  [arXiv:1110.3327 [astro-ph.CO]].
  
\bibitem{Domcke:2016bkh}
  V.~Domcke, M.~Pieroni and P.~Bin�truy,
  JCAP {\bf 1606} (2016) 031
  [arXiv:1603.01287 [astro-ph.CO]].
  
\bibitem{Bartolo:2016ami}
  N.~Bartolo {\it et al.},
  JCAP {\bf 1612} (2016) no.12,  026
  [arXiv:1610.06481 [astro-ph.CO]].

  
  
\bibitem{Endlich:2012pz}
  S.~Endlich, A.~Nicolis and J.~Wang,
  JCAP {\bf 1310} (2013) 011
  [arXiv:1210.0569 [hep-th]].


\bibitem{Gruzinov:2004ty}
  A.~Gruzinov,
  Phys.\ Rev.\ D {\bf 70} (2004) 063518
  [astro-ph/0404548].

\bibitem{Nicolis:2013lma}
  A.~Nicolis, R.~Penco and R.~A.~Rosen,
  Phys.\ Rev.\ D {\bf 89} (2014) no.4,  045002
  [arXiv:1307.0517 [hep-th]].

\bibitem{Bartolo:2015qvr}
  N.~Bartolo, D.~Cannone, A.~Ricciardone and G.~Tasinato,
  JCAP {\bf 1603} (2016) no.03,  044
  [arXiv:1511.07414 [astro-ph.CO]].


\bibitem{Cannone:2014uqa}
  D.~Cannone, G.~Tasinato and D.~Wands,
  JCAP {\bf 1501} (2015) no.01,  029
  [arXiv:1409.6568 [astro-ph.CO]].

\bibitem{Higuchi:1986py}
  A.~Higuchi,
  Nucl.\ Phys.\ B {\bf 282} (1987) 397.



\bibitem{Fasiello:2013woa} 
  M.~Fasiello and A.~J.~Tolley,
  JCAP {\bf 1312}, 002 (2013)
  [arXiv:1308.1647 [hep-th]].


\bibitem{Ricciardone:2016lym}
  A.~Ricciardone and G.~Tasinato,
  Phys.\ Rev.\ D {\bf 96} (2017) no.2,  023508
  [arXiv:1611.04516 [astro-ph.CO]].


\bibitem{Maldacena:2011nz}
  J.~M.~Maldacena and G.~L.~Pimentel,
  JHEP {\bf 1109} (2011) 045
  [arXiv:1104.2846 [hep-th]].
  
\bibitem{Arnowitt:1962hi}
  R.~L.~Arnowitt, S.~Deser and C.~W.~Misner,
  Gen.\ Rel.\ Grav.\  {\bf 40} (2008) 1997
  [gr-qc/0405109].
  


\bibitem{Cheung:2007st}
  C.~Cheung, P.~Creminelli, A.~L.~Fitzpatrick, J.~Kaplan and L.~Senatore,
  JHEP {\bf 0803} (2008) 014
  [arXiv:0709.0293 [hep-th]].


\bibitem{Horndeski:1974wa}
  G.~W.~Horndeski,
  Int.\ J.\ Theor.\ Phys.\  {\bf 10} (1974) 363.


\bibitem{Kobayashi:2010cm}
  T.~Kobayashi, M.~Yamaguchi and J.~Yokoyama,
  Phys.\ Rev.\ Lett.\  {\bf 105} (2010) 231302
  [arXiv:1008.0603 [hep-th]];
  
  T.~Kobayashi, M.~Yamaguchi and J.~Yokoyama,
  Prog.\ Theor.\ Phys.\  {\bf 126} (2011) 511
  [arXiv:1105.5723 [hep-th]].

\bibitem{Germani:2010gm}
  C.~Germani and A.~Kehagias,
  Phys.\ Rev.\ Lett.\  {\bf 105} (2010) 011302
  [arXiv:1003.2635 [hep-ph]];
  
  C.~Germani and A.~Kehagias,
  JCAP {\bf 1005} (2010) 019
   Erratum: [JCAP {\bf 1006} (2010) E01]
  [arXiv:1003.4285 [astro-ph.CO]].

\bibitem{Ade:2015hxq}
  P.~A.~R.~Ade {\it et al.} [Planck Collaboration],
  Astron.\ Astrophys.\  {\bf 594} (2016) A16
  [arXiv:1506.07135 [astro-ph.CO]].


\bibitem{Domenech:2017kno}
  G.~Dom�nech, T.~Hiramatsu, C.~Lin, M.~Sasaki, M.~Shiraishi and Y.~Wang,
  JCAP {\bf 1705} (2017) no.05,  034
  [arXiv:1701.05554 [astro-ph.CO]].


\bibitem{Burrage:2010cu}
  C.~Burrage, C.~de Rham, D.~Seery and A.~J.~Tolley,
  JCAP {\bf 1101} (2011) 014
  [arXiv:1009.2497 [hep-th]].

\bibitem{Copeland:2012qf}
  E.~J.~Copeland, A.~Padilla and P.~M.~Saffin,
  JCAP {\bf 1212} (2012) 026
  [arXiv:1208.3373 [hep-th]].

\bibitem{Mukhanov:1990me}
  V.~F.~Mukhanov, H.~A.~Feldman and R.~H.~Brandenberger,
  Phys.\ Rept.\  {\bf 215} (1992) 203.


\bibitem{Maldacena:2002vr}
  J.~M.~Maldacena,
  JHEP {\bf 0305} (2003) 013
  [astro-ph/0210603].


\bibitem{Riotto:2002yw}
  A.~Riotto,
  ICTP Lect.\ Notes Ser.\  {\bf 14} (2003) 317
  [hep-ph/0210162].


\bibitem{Creminelli:2014wna}
  P.~Creminelli, J.~Gleyzes, J.~Nore�a and F.~Vernizzi,
  Phys.\ Rev.\ Lett.\  {\bf 113} (2014) no.23,  231301
  [arXiv:1407.8439 [astro-ph.CO]].


\bibitem{Baumann:2015xxa}
  D.~Baumann, H.~Lee and G.~L.~Pimentel,
  JHEP {\bf 1601} (2016) 101
  [arXiv:1507.07250 [hep-th]].

\bibitem{Fumagalli:2016afy}
  J.~Fumagalli, S.~Mooij and M.~Postma,
  arXiv:1610.08460 [gr-qc].


\bibitem{Bordin:2016ruc}
  L.~Bordin, P.~Creminelli, M.~Mirbabayi and J.~Nore�a,
  JCAP {\bf 1609} (2016) no.09,  041
  [arXiv:1605.08424 [astro-ph.CO]].



\bibitem{Dimastrogiovanni:2014ina}
  E.~Dimastrogiovanni, M.~Fasiello, D.~Jeong and M.~Kamionkowski,
  JCAP {\bf 1412} (2014) 050
  [arXiv:1407.8204 [astro-ph.CO]].


\bibitem{Meerburg:2016ecv}
  P.~D.~Meerburg, J.~Meyers, A.~van Engelen and Y.~Ali-Ha�moud,
  Phys.\ Rev.\ D {\bf 93} (2016) 123511
  [arXiv:1603.02243 [astro-ph.CO]].


\bibitem{Endlich:2013jia}
  S.~Endlich, B.~Horn, A.~Nicolis and J.~Wang,
  Phys.\ Rev.\ D {\bf 90} (2014) no.6,  063506
  [arXiv:1307.8114 [hep-th]].


\bibitem{Akhshik:2014bla}
  M.~Akhshik,
  JCAP {\bf 1505} (2015) no.05,  043
  [arXiv:1409.3004 [astro-ph.CO]].


\bibitem{Biagetti:2013kwa}
  M.~Biagetti, M.~Fasiello and A.~Riotto,
  Phys.\ Rev.\ D {\bf 88} (2013) 103518
  [arXiv:1305.7241 [astro-ph.CO]].


\bibitem{Biagetti:2017viz}
  M.~Biagetti, E.~Dimastrogiovanni and M.~Fasiello,
  arXiv:1708.01587 [astro-ph.CO].


\bibitem{solphen}
  N.~Bartolo, S.~Matarrese, M.~Peloso and A.~Ricciardone,
  JCAP {\bf 1308} (2013) 022
  [arXiv:1306.4160 [astro-ph.CO]];

  N.~Bartolo, M.~Peloso, A.~Ricciardone and C.~Unal,
  JCAP {\bf 1411} (2014) no.11,  009
  [arXiv:1407.8053 [astro-ph.CO]];
  M.~Akhshik, R.~Emami, H.~Firouzjahi and Y.~Wang,
  JCAP {\bf 1409} (2014) 012
  [arXiv:1405.4179 [astro-ph.CO]];
  C.~Lin and L.~Z.~Labun,
  JHEP {\bf 1603} (2016) 128
  [arXiv:1501.07160 [hep-th]];
  D.~Cannone, J.~O.~Gong and G.~Tasinato,
  JCAP {\bf 1508} (2015) no.08,  003
  [arXiv:1505.05773 [hep-th]];
  A.~A.~Abolhasani, M.~Akhshik, R.~Emami and H.~Firouzjahi,
  JCAP {\bf 1603} (2016) 020
  [arXiv:1511.03218 [astro-ph.CO]];
  T.~Rostami, A.~Karami and H.~Firouzjahi,
  JCAP {\bf 1706} (2017) no.06,  039
  [arXiv:1702.03744 [astro-ph.CO]].




\bibitem{AmaroSeoane:2012km}
  P.~Amaro-Seoane {\it et al.},
  GW Notes {\bf 6} (2013) 4
  [arXiv:1201.3621 [astro-ph.CO]];


\bibitem{earlydirect}
  A.~R.~Liddle,
  Phys.\ Rev.\ D {\bf 49} (1994) 3805
   Erratum: [Phys.\ Rev.\ D {\bf 51} (1995) 4603]
  [gr-qc/9307036];
  R.~Bar-Kana,
  Phys.\ Rev.\ D {\bf 50} (1994) 1157
  [astro-ph/9401050];

  Phys.\ Rev.\ D {\bf 55} (1997) R435
  [astro-ph/9607066];
  T.~L.~Smith, M.~Kamionkowski and A.~Cooray,
  Phys.\ Rev.\ D {\bf 73} (2006) 023504
  [astro-ph/0506422].



\bibitem{Lyth:2009zz}
  D.~H.~Lyth and A.~R.~Liddle,
  Cambridge, UK: Cambridge Univ. Pr. (2009) 497 p

\bibitem{Harry:2010zz}
  G.~M.~Harry [LIGO Scientific Collaboration],
  Class.\ Quant.\ Grav.\  {\bf 27} (2010) 084006.

\bibitem{Kuroyanagi:2014nba}
  S.~Kuroyanagi, T.~Takahashi and S.~Yokoyama,
  JCAP {\bf 1502} (2015) 003
  [arXiv:1407.4785 [astro-ph.CO]].


\bibitem{Gao:2011vs}
  X.~Gao, T.~Kobayashi, M.~Yamaguchi and J.~Yokoyama,
  Phys.\ Rev.\ Lett.\  {\bf 107} (2011) 211301
  [arXiv:1108.3513 [astro-ph.CO]].



  
\bibitem{tenslongA}
  K.~W.~Masui and U.~L.~Pen,
  Phys.\ Rev.\ Lett.\  {\bf 105} (2010) 161302
  [arXiv:1006.4181 [astro-ph.CO]];
  D.~Jeong and M.~Kamionkowski,
  Phys.\ Rev.\ Lett.\  {\bf 108} (2012) 251301
  [arXiv:1203.0302 [astro-ph.CO]];
  L.~Dai, D.~Jeong and M.~Kamionkowski,
  Phys.\ Rev.\ D {\bf 88} (2013) no.4,  043507
  [arXiv:1306.3985 [astro-ph.CO]];
  L.~Dai, D.~Jeong and M.~Kamionkowski,
  Phys.\ Rev.\ D {\bf 87} (2013) no.10,  103006
  [arXiv:1302.1868 [astro-ph.CO]];
  S.~Brahma, E.~Nelson and S.~Shandera,
  Phys.\ Rev.\ D {\bf 89} (2014) no.2,  023507
  [arXiv:1310.0471 [astro-ph.CO]].
  
  
\bibitem{Barnaby:2010vf}
  N.~Barnaby and M.~Peloso,
  Phys.\ Rev.\ Lett.\  {\bf 106} (2011) 181301
  [arXiv:1011.1500 [hep-ph]].
  
\bibitem{Barnaby:2011vw}
  N.~Barnaby, R.~Namba and M.~Peloso,
  JCAP {\bf 1104} (2011) 009
  [arXiv:1102.4333 [astro-ph.CO]].
  

\bibitem{Soda:2011am}
  J.~Soda, H.~Kodama and M.~Nozawa,
  JHEP {\bf 1108} (2011) 067
  [arXiv:1106.3228 [hep-th]];
  M.~Shiraishi, D.~Nitta and S.~Yokoyama,
  Prog.\ Theor.\ Phys.\  {\bf 126} (2011) 937
  [arXiv:1108.0175 [astro-ph.CO]].


  
\bibitem{Shiraishi:2013kxa}
  M.~Shiraishi, A.~Ricciardone and S.~Saga,
  JCAP {\bf 1311} (2013) 051
  [arXiv:1308.6769 [astro-ph.CO]].

  
  
  
  \bibitem{inNGa}
  D.~Hanson and A.~Lewis,
  Phys.\ Rev.\ D {\bf 80} (2009) 063004
  [arXiv:0908.0963 [astro-ph.CO]];
  S.~B.~Giddings and M.~S.~Sloth,
  JCAP {\bf 1101} (2011) 023
  [arXiv:1005.1056 [hep-th]];
  S.~B.~Giddings and M.~S.~Sloth,
  Phys.\ Rev.\ D {\bf 84} (2011) 063528
  [arXiv:1104.0002 [hep-th]];
  S.~Nurmi, C.~T.~Byrnes and G.~Tasinato,
  JCAP {\bf 1306} (2013) 004
  [arXiv:1301.3128 [astro-ph.CO]];
  C.~T.~Byrnes, S.~Nurmi, G.~Tasinato and D.~Wands,
  JCAP {\bf 1203} (2012) 012
  [arXiv:1111.2721 [astro-ph.CO]];
  M.~Gerstenlauer, A.~Hebecker and G.~Tasinato,
  JCAP {\bf 1106} (2011) 021
  [arXiv:1102.0560 [astro-ph.CO]];
  C.~T.~Byrnes, M.~Gerstenlauer, S.~Nurmi, G.~Tasinato and D.~Wands,
  JCAP {\bf 1010} (2010) 004
  [arXiv:1007.4277 [astro-ph.CO]];
  C.~T.~Byrnes, M.~Gerstenlauer, A.~Hebecker, S.~Nurmi and G.~Tasinato,
  JCAP {\bf 1008} (2010) 006
  [arXiv:1005.3307 [hep-th]];
  E.~Nelson and S.~Shandera,
  Phys.\ Rev.\ Lett.\  {\bf 110} (2013) no.13,  131301
  [arXiv:1212.4550 [astro-ph.CO]];
  M.~LoVerde, E.~Nelson and S.~Shandera,
  JCAP {\bf 1306} (2013) 024
  [arXiv:1303.3549 [astro-ph.CO]].
  S.~Adhikari, S.~Shandera and A.~L.~Erickcek,
  Phys.\ Rev.\ D {\bf 93} (2016) no.2,  023524
  [arXiv:1508.06489 [astro-ph.CO]].


\bibitem{Cornish:2001bb}
  N.~J.~Cornish,
  Phys.\ Rev.\ D {\bf 65} (2002) 022004
  [gr-qc/0106058].



\bibitem{Kato:2015bye}
  R.~Kato and J.~Soda,
  Phys.\ Rev.\ D {\bf 93} (2016) no.6,  062003
  [arXiv:1512.09139 [gr-qc]].

\bibitem{Romano:2016dpx}
  J.~D.~Romano and N.~J.~Cornish,
  Living Rev.\ Rel.\  {\bf 20} (2017) 2
  [arXiv:1608.06889 [gr-qc]].


\bibitem{Smith:2016jqs}
  T.~L.~Smith and R.~Caldwell,
  Phys.\ Rev.\ D {\bf 95} (2017) no.4,  044036
  [arXiv:1609.05901 [gr-qc]].

\bibitem{Thorne:2017jft}
  B.~Thorne, T.~Fujita, M.~Hazumi, N.~Katayama, E.~Komatsu and M.~Shiraishi,
  arXiv:1707.03240 [astro-ph.CO].

\bibitem{Allen:1996vm}
  B.~Allen,
  In *Les Houches 1995, Relativistic gravitation and gravitational radiation* 373-417
  [gr-qc/9604033].


\end{thebibliography}
\end{document}